 \documentstyle[prl,aps,epsfig]{revtex}

\begin{document}
 \renewcommand{\theequation}{\thesection.\arabic{equation}}

 \title{On the Space-Time Symmetries of Non-Commutative Gauge Theories}
 \author{A. Iorio$^{a,b,}$\thanks{E-mail: iorio@sa.infn.it},
     and T. S\'{y}kora$^{a,c,}$\thanks{E-mail: tomas.sykora@mff.cuni.cz} }
 \address{$^a$ Center for Theoretical Physics,
 Massachusetts Institute of Technology, Cambridge, MA, 02139-4307 USA}
 \address{$^b$Dipartimento di Fisica "E.R. Caianiello"
  Universit\`a di Salerno, 84081 Baronissi (SA), Italy, and I.N.F.N.}
\address{$^c$ Institute of Particle and Nuclear Physics, Charles University Prague,
V Hole\v{s}ovi\v{c}k\'{a}ch 2, 182 00 Prague, Czech Republic}

 \date{\today}
 \maketitle
\begin{abstract}
We study the  space-time symmetries and transformation properties
of the non-commutative U(1) gauge theory, by using Noether
charges. We carry out our analysis by keeping an open view on the
possible ways $\theta^{\mu \nu}$ could transform. We conclude that
$\theta^{\mu \nu}$ cannot transform under any space-time
transformation since the theory is not invariant under the
conformal transformations, with the only exception of space-time
translations. The same analysis applies to other gauge groups.
\end{abstract}

\bigskip
\bigskip


\noindent Keyword(s): gauge field theories, non-commutative
geometry, symmetry and conservation laws.



\noindent MIT-CTP-3206


\section{Introduction}
The idea of non-commuting space-time coordinates was first
considered by Heisenberg and written about by Snyder
\cite{Snyder}. The geometry was widely developed in recent years
by Connes \cite{Connes}. Following the discovery of Seiberg and
Witten \cite{SW} of a map (SW map) that relates non-commutative to
commutative gauge theories, there has been an increasing interest
in studying non-commutativity from a theoretical point of view, as
well as its impact on phenomenology \cite{Jac1}, \cite{Cai}. Our
purpose is to study the space-time symmetry properties of the U(1)
non-commutative gauge theory. The extension of our analysis to
other gauge groups is straightforward.

The most common explicit realizations of the non-commutative nature of
space-time coordinates are  \cite{Wess}: the canonical structure, the Lie
algebra structure, and the $q$-deformed space structure. In all cases the algebraic structure
is associative.

We shall work with the canonical structure, given by
\begin{equation} \label{1}
x^\mu * x^\nu - x^\nu * x^\mu
= i \theta^{\mu \nu} \;,
\end{equation}
where the Moyal-Weyl $*$-product of two fields $\phi(x)$ and $\chi(x)$ is defined as
\begin{eqnarray}
(\phi  * \chi) (x) & \equiv &
\exp\{ \frac{i}{2}  \theta^{\mu \nu} \partial^x_\mu \partial^y_\nu \} \phi (x) \chi (y)|_{y \to x}  \nonumber \\
& = &  \phi (x) \chi (x)
+ \frac{i}{2}  \theta^{\mu \nu} \partial_\mu \phi (x) \partial_\nu \chi (x)
- \frac{1}{8} \theta^{\mu \nu} \theta^{\lambda \kappa} \partial_\mu \partial_\lambda \phi (x) \partial_\nu \partial_\kappa \chi (x)
+ \cdots  \;, \label{3}
\end{eqnarray}
$\theta^{\mu \nu}$ is $c$-number valued, the Greek indices run from $0$ to $n-1$, and $n$
is the dimension of the space-time.

The action for the non-commutative Abelian gauge theory in four dimensions is
\begin{equation}\label{4}
\hat{I} = - \frac{1}{4} \int d^4 x \hat{F}^{\mu \nu} \hat{F}_{\mu \nu} \;,
\end{equation}
where $ \hat{F}_{\mu \nu} = \partial_\mu \hat{A}_\nu -
\partial_\nu \hat{A}_\mu - i [ \hat{A}_\mu , \hat{A}_\nu ]_* $,
$\hat{A}_\mu$ can be expressed in terms of a U(1) gauge field
$A_\mu$ and of $\theta^{\mu \nu}$ by means of the SW map,
$\hat{A}_\mu (A,\theta)$. Note that  $\hat{A}_\mu (A,\theta) \to
A_\mu$ as $\theta^{\mu \nu} \to 0$, hence, in that limit,
$\hat{F}_{\mu \nu} \to F_{\mu \nu} =
\partial_\mu A_\nu - \partial_\nu A_\mu$.

The action (\ref{4}) arises  either as the non-commutative analog
of a standard gauge theory or as the low energy limit of some
string theory \cite{SW}. In both cases the physical nature of the
gauge field $A_\mu$ and of $\theta^{\mu \nu}$ is not clear, and
has to be investigated. For instance, one sees that within the
$C^*$ algebra defined by (\ref{1}), the $*$-multiplication by
$x_\mu$ (thought as an adjoint action) is a derivation
$\partial_\mu$. It is this derivative that one makes covariant by
introducing the field $\hat{A}_\mu$ \cite{Wess}:
\begin{equation} \label{xA}
x^\mu \to x^\mu + \theta^{\mu \nu} \hat{A}_\nu
\end{equation}
while, in the standard case, the gauge prescription acts on the
momentum $p_\mu$: $p_\mu \to p_\mu +  A_\mu$. Thus, even if the
formal resemblance entitles us to treat $\hat{A}_\mu$ as the
four-potential of electrodynamics \cite{Jac1}, \cite{Cai}, one has
to perform an open minded investigation of the properties of the
theory (\ref{4}). An issue is the nature of $\theta^{\mu \nu}$.
Some recent investigations have considered $\theta^{\mu \nu}$ as a
second rank tensor under Lorentz transformations, with scaling
properties imposed by demanding the action to be scale invariant
\cite{7A}. In this paper we would like to address the problem of
the space-time symmetries and transformation properties of the
theory (\ref{4}) from a dynamical perspective, namely by using the
Noether charges. This will allow us to have stronger constraints
on the possible ways $\theta^{\mu \nu}$ can transform.

\section{Dynamically Consistent Transformations}

There are various ways to represent the conformal algebra as transformations acting on fields.
An example are the  the so-called geometric transformations, namely the effect of the infinitesimal
variation of the coordinates
\begin{equation}
{x'}^\mu - x^\mu = \delta_f x^\mu \equiv -f^\mu (x) \;,
\end{equation}
on the fields, evaluated at the same point $x$ (for a review see
e.g.  \cite{Jac2}). Since we shall be dealing with the action
(\ref{4}) we are interested in the transformations of the gauge
field and the field strength
\begin{equation} \label{5}
\delta_f A_\mu \equiv A'_\mu (x) - A_\mu (x) \quad {\rm and} \quad
\delta_f F_{\mu \nu} \equiv F'_{\mu \nu} (x) - F_{\mu \nu} (x) \;,
\end{equation}
A separate issue are the transformations of $\theta^{\mu \nu}$, on which we shall concentrate in the next Section.

The infinitesimal quantities $f^\mu$ take the following form
\begin{equation} \label{6}
f^\mu = a^\mu \quad {\rm or} \quad
f^\mu = \omega^\mu_\nu x^\nu \quad {\rm or} \quad
f^\mu = a x^\mu  \quad {\rm or} \quad
f^\mu = a^\mu x^2 - 2 a \cdot x x^\mu \;,
\end{equation}
for infinitesimal translations, rotations (and boosts), dilations,
and special conformal transformations, respectively, where, as
usual, $\omega^{\mu \nu} = - \omega^{\nu \mu}$.

It can be easily shown that the geometric transformations (\ref{5}) can be written as
\begin{equation} \label{7}
\delta_f A_\mu = {\bf L}_f A_\mu \quad {\rm and} \quad
\delta_f F_{\mu \nu} = {\bf L}_f F_{\mu \nu} =
\partial_\mu {\bf L}_f A_\nu -
\partial_\nu {\bf L}_f A_\mu \;,
\end{equation}
where ${\bf L}_f$ is the standard {\it Lie derivative}
\begin{equation} \label{lie}
{\bf L}_f X_{\mu \dots \nu}^{\lambda \dots \kappa} =
f^\alpha \partial_\alpha X_{\mu \dots \nu}^{\lambda \dots \kappa}
+ (\partial_\mu f^\alpha) X_{\alpha \dots \nu}^{\lambda \dots \kappa}
+ \cdots
+ (\partial_\nu f^\alpha) X_{\mu \dots \alpha}^{\lambda \dots \kappa}
- (\partial_\alpha f^\lambda) X_{\mu \dots \nu}^{\alpha \dots \kappa}
- \cdots
- (\partial_\alpha f^\kappa) X_{\mu \dots \nu}^{\lambda \dots \alpha} \;.
\end{equation}
Note also that one can write the identity
\begin{equation} \label{8a}
\delta_f A_\mu = {\bf L}_f A_\mu \equiv f^{\alpha} F_{\alpha\mu}+
D_{\mu}(f^{\alpha}A_{\alpha})\;,
\end{equation}
where $D_\mu = \partial_\mu - i A_\mu$ is the covariant
derivative. Since
\begin{equation} \label{8}
[ \delta_f , \delta_g ] A_\mu = \delta_{[f,g]} A_\mu \quad {\rm and} \quad
[ \delta_f , \delta_g ] F_{\mu \nu} = \delta_{[f,g]} F_{\mu \nu} \;,
\end{equation}
one sees that indeed this is a representation of the conformal
algebra.

On the other hand, one can also introduce the {\it covariant}
geometric transformations  \cite{Jac3}, namely the transformations
$\bar{\delta}_f$ obtained by gauge-transforming $\delta_f$. For
instance the transformation of the gauge field given in (\ref{8a})
becomes
\begin{equation} \label{9a}
\bar{\delta_f} A_\mu = f^{\alpha} F_{\alpha\mu} \;.
\end{equation}
This time one has
\begin{equation} \label{9}
[ \bar{\delta}_f , \bar{\delta}_g ] A_\mu = \bar{\delta}_{[f,g]} A_\mu + D_\mu X \quad {\rm and} \quad
[ \bar{\delta}_f , \bar{\delta}_g ] F_{\mu \nu} = \bar{\delta}_{[f,g]} F_{\mu \nu} + D_\mu X_\nu \;,
\end{equation}
thus, as a representation of the original algebra, it closes only
up to a gauge transformation, that one is free to perform. For a
generalization of the latter to the non-commutative counterparts
see \cite{JacPi}.

In these two examples there is no need to refer to an action to
obtain the way the fields have to transform. In this sense there
is a certain amount of arbitrariness, and there is no need for
these transformations to be symmetries. For instance, one can
easily check that the scaling properties of fields of arbitrary
spin obtained via geometric transformations (either $\delta_{f}$
or $\bar{\delta}_{f}$) do not always agree with the standard
 scaling properties. Note that this is not the case for $A_\mu$ and $F_{\mu \nu}$ in four dimensions,
where the geometric scaling properties agree with the standard ones.

Another possible way to represent the conformal algebra acting on the fields is via Noether charges.
This (infinite dimensional, reducible) representation has a very important role among the various
representations because is the form of the action that dictates the transformations of the fields,
whether or not they are symmetries. We shall call these transformations $\Delta_f$.

In a theory with Lagrangian density ${\cal L} (\Phi_i , \partial \Phi_i)$ (where the collective
index $i$ takes care of the different fields, as well as their spin type) the Noether current for
space-time transformations has the form
\begin{equation} \label{10}
J^\mu_f = \Pi^{\mu i} \delta_f \Phi_i + {\cal L} \delta_f x^\mu \;,
\end{equation}
where
$
\Pi^{\mu i} = \delta {\cal L} / \delta \partial_\mu \Phi_i
$, and, as in (\ref{5}),
$
\delta_f \Phi_i = {\Phi'}_i (x) - \Phi_i (x)
$.
With the current (\ref{10}) one can:
i)test whether the given transformation is a symmetry by picking the correspondent $f^\mu$ in (\ref{6}), and
checking whether $\partial_\mu J^\mu = 0$, by using the equations of motion;
ii) use the Noether charges
$
Q_f \equiv \int d^3 x J_f^0
$,
and the canonical equal-time Poisson brackets
$
\{ \Phi_i (x) ,  \Pi^j (y) \} = \delta^j_i \delta (\vec{x} - \vec{y})
$,
to generate the transformations of an arbitrary function of the canonical variables
\begin{equation} \label{11}
\{ G (\Phi_i , \Pi_i) , Q_f \} \equiv \Delta_f G (\Phi_i , \Pi_i) \;.
\end{equation}
Note also that, for $f^{0} = g^{0}=0$,
\begin{equation} \label{12}
\{ Q_f , Q_g \} = Q_{[f,g]} \;,
\end{equation}
and Eq.s (\ref{11}) and  (\ref{12}) hold whether or not $\partial_0 Q_f = 0$.

Of course, when $Q_f$ acts on the fields it must reproduce the transformations
one started with $\Delta_f \Phi_i = \delta_f \Phi_i$. We call $\Delta_f$ the
{\it dynamically consistent transformations}.

The expression for the current (\ref{10}) has been obtained by
varying the action, including the measure, under an arbitrary
space-time transformation, and only afterwards one tests the
invariance, following the above described procedure. Sometimes the
Noether theorem is used in a somehow different perspective, namely
by checking that $\delta {\cal L} = \partial_\mu V^\mu$, which is
only true for invariant actions, and then writing the current as
$J^\mu = \Pi^{\mu i} \delta^{*} \Phi_i - V^\mu$, where in this
case $\delta^{*} \Phi_i = {\Phi'}_i (x') - \Phi_i (x)$, hence
$\delta^{*} \neq \delta_f$. The choice $\delta^{*}$ is the most
useful in the case of gauge transformations and supersymmetry
\cite{Io2}, while for standard space-time transformations, as the
ones we are considering in this paper, it is more convenient to
use (\ref{10}). Of course, the two procedures are consistent
\cite{Io2}, \cite{Lop}.

\section{Non-Commutative U(1) Gauge Theory}

For sake of clarity, let us now write down the terms in the action in
(\ref{4}) up to first order in $\theta$
\begin{equation}\label{13}
\hat{I} = - \frac{1}{4} \int  d^4 x \; [F^{\mu \nu} F_{\mu \nu}
      -\frac{1}{2} \theta^{\alpha \beta} F_{\alpha \beta} F^{\mu \nu} F_{\mu \nu}
      + 2 \theta^{\alpha \beta} F_{\alpha \mu} F_{\beta \nu} F^{\mu \nu}] + O(\theta^2) \;,
\end{equation}
although the following results hold to  any order in $\theta$. The
$\theta$-dependence in (\ref{13}) is obtained from the $*$-product
in $\hat{F}_{\mu \nu}$, and by using the SW map
\begin{equation}
\hat{A}_{\mu}(A, \theta)=A_{\mu}-\frac{1}{2}\theta^{\alpha
\beta}A_{\alpha}(\partial_{\beta}A_{\mu} + F_{\beta \mu}) +
O(\theta^{2}) \;,
\end{equation}
which solves the Seiberg-Witten equation \cite{SW}
\begin{equation}
\hat{A}(A+\delta_{\lambda} A)=\hat{A}(A)+\hat{\delta}_{\hat{\lambda}}\hat{A}(A)\;,
\end{equation}
where $\delta_{\lambda}A_{\mu}=\partial_{\mu}\lambda$,
$\hat{\delta}_{\hat{\lambda}}\hat{A}_{\mu}=\partial_{\mu}\hat{\lambda}
-i [ \hat{A}_{\mu} , \hat{\lambda}]_*$, and $\lambda$,
$\hat{\lambda}(\lambda,A)$ are the parameters of the gauge
transformations for standard and non-commutative U(1) gauge group,
respectively.

The Noether current for space-time transformations that we obtain from the action (\ref{13}) is then
\begin{equation} \label{15}
J_f^\mu = \Pi^{\mu \nu} \delta_f A_\nu - {\cal L} f^\mu \;,
\end{equation}
where the $f^\mu$'s are given in (\ref{6}), $\Pi^{\mu \nu} =
\delta {\cal L} / \delta \partial_\mu A_\nu$, and, being
$\Pi^\mu_{\alpha \beta} = \delta {\cal L} / \delta \partial_\mu
\theta^{\alpha \beta} = 0$, the transformations $\delta_f
\theta^{\alpha \beta}$ do not enter the Noether current.

Let us now analyze the symmetry properties by writing the divergence of this current as
\begin{equation}\label{16}
\partial_\mu J_f^\mu = \Pi^{\mu \nu} F_{\alpha \nu} \partial_\mu f^\alpha - {\cal L} \partial_\mu f^\mu \;.
\end{equation}
We obtain that $\partial_\mu J_f^\mu = 0$ for infinitesimal translations $f^\mu = a^\mu$, but
\begin{equation} \label{18}
\partial_\mu J_f^\mu = \omega^\alpha_\mu \Pi^{\mu \nu} F_{\alpha \nu} \;,
\quad \quad
\partial_\mu J_f^\mu = a (\Pi^{\mu \nu} F_{\mu \nu} - 4 {\cal L}) \;,
\quad \quad
\partial_\mu J_f^\mu = 2 \Pi^{\mu \nu} F_{\alpha \nu} (a^\alpha x_\mu - a_\mu x^\alpha - a \cdot x \delta^\alpha_\mu)
+ 8 a \cdot x {\cal L} \;,
\end{equation}
for infinitesimal Lorentz transformations, dilations and special
conformal transformations, respectively. In absence of the
non-commutative corrections, $\theta^{\mu \nu} = 0$, $\Pi^{\mu
\nu} = - F^{\mu \nu}$, and from (\ref{18}) one immediately sees
that $\partial_\mu J_f^\mu = 0$ for all $f^\mu$'s in (\ref{6}),
and the theory is invariant under the full conformal
transformations\footnote{Since $A_\mu$ is a spin 1 field, and we
are in four dimensions, for $\theta^{\mu \nu} = 0$, scale
invariance alone implies special conformal invariance
\cite{Io1}.}.

On the other hand, the action $\hat I$, for $\theta^{\mu \nu} \neq
0$, is only invariant under translations. Of course, this leaves
room to special choices of the parameters and/or of the field
configurations, to obtain conserved currents. For instance, if one
performs two {\it dependent} infinitesimal boosts with parameters
$\omega^{0}_1 = \omega^{0}_2 = \omega$, from the first condition
in (\ref{18}) one obtains $\Pi^{12} (E_2 - E_1) = E_3 (\Pi^{13} +
\Pi^{23})$. Thus, for instance, one finds conservation if the
electric field $\vec{E}$ lives in the $(1,2)$-plane, and has equal
components.

It is crucial to notice that the results in (\ref{18}) hold for
the full theory (\ref{4}), to all orders in $\theta$. As a matter
of fact, to obtain (\ref{18}) we have only used the conditions
\begin{equation} \label{17}
\partial_\mu \theta = 0 \;, \quad  \quad \Pi^{\mu \nu} = - \Pi^{\nu \mu} \;,\quad  \quad
\partial_\mu \Pi^{\mu \nu} = 0 \;.
\end{equation}
i.e. i) $\theta^{\mu \nu}$ does not depend on the coordinates
$x^\mu$; ii) $\Pi^{\mu \nu}$ is antisymmetric, which is a
consequence of the antisymmetry of $F_{\mu \nu}$; iii) the
equations of motion for $A_\mu$ are satisfied in the form
$\partial_\mu \Pi^{\mu \nu} = 0$, where the latter expression is
valid to all orders in $\theta$ provided one changes $\Pi^{\mu
\nu}$ accordingly. Furthermore $\Pi^\mu_{\alpha\beta} = 0$ to all
orders.

We now want to check the dynamical consistency of the
transformations, along the lines of what explained in Section II.
It is straightforward to see that for $A_\mu$ and $F_{\mu \nu}$
\begin{equation} \label{19}
\Delta_f A_\mu = \delta_f A_\mu \quad {\rm and} \quad \Delta_f
F_{\mu \nu} = \delta_f F_{\mu \nu} \;,
\end{equation}
while for $\theta^{\mu \nu}$
\begin{equation} \label{20}
\Delta_f \theta^{\mu \nu} = 0 \;,
\end{equation}
for all $f^\mu$'s in (\ref{6}).

At this point one has to investigate whether $\theta^{\mu \nu}$
could be treated as a dummy field $D$, in the spirit of what
happens in supersymmetric models. The answer in no. In both cases
$\delta_f \theta^{\mu \nu}$ and $\delta_{\rm SUSY} D$ do not
explicitly enter the expression of the  Noether current, hence one
can arbitrarily choose them. Nonetheless, in supersymmetric
models: i) one fixes $\delta_{\rm SUSY} D$ by demanding the action
$I_{\rm SUSY}$ (which contains the dummy fields explicitly) to be
invariant; ii) even if $\Pi_D = 0$, $\Delta_{\rm SUSY} D \neq 0$
due to the fact that $D$ can be properly expressed in terms of the
dynamical fields by means of the constraint $\delta I_{\rm SUSY} /
\delta D = 0$ \cite{Io2}. Therefore $\delta_{\rm SUSY}$ represents
the SUSY algebra {\it off-shell}, while $\Delta_{\rm SUSY}$
represents the SUSY algebra {\it on-shell}, and both are
symmetries of the action. On the contrary, in our case, if one
uses
\begin{equation} \label{tom}
\frac{\delta {\hat I}}{\delta \theta^{\mu \nu}} = 0 \;,
\end{equation}
as a constraint, then one obtains different dynamical
transformations of $\theta^{\mu \nu}$ at different orders, and
trivial theories. For instance, for the first order action
(\ref{13}), one cannot express $\theta^{\mu \nu}$ in terms of the
dynamical fields, hence there is no contribution to $\Delta_f
\theta^{\mu \nu}$ from $\Delta_f A_\mu$ or $\Delta_f F_{\mu \nu}$.
While, at higher orders, there are not simple expressions for
$\theta^{\mu \nu}$ in terms of $A_\mu$ or $F_{\mu \nu}$, and
$\Delta_f \theta^{\mu \nu}$ has different expressions for
different orders. Furthermore, the constraints imposed on $F_{\mu
\nu}$ by (\ref{tom}) make the theory trivial. For instance, at
first order, (\ref{tom}) implies $F_{\mu \nu} = 0$, which is too
trivial a theory. Note also that if one uses (\ref{tom}) the
expressions of $\partial_\mu J^\mu_f$ in (\ref{18}) are trivially
zero for all conformal transformations.

We conclude that $\delta_f \theta^{\mu \nu}$ cannot be fixed by
any symmetry requirement (with the only exception of
translations), and that to have a physical meaningful theory one
should not make use of the "equations of motion" (\ref{tom}).
Therefore, among all possible $\delta_f \theta^{\mu \nu}$'s that
represent the conformal algebra, the most natural choice seems to
us $\delta_f \theta^{\mu \nu} = \Delta_f \theta^{\mu \nu} = 0$
(which agrees with the translation invariance), and $\theta^{\mu
\nu}$ does not transform under dynamically consistent space-time
transformations. This choice is also consistent with
considerations in \cite{JacPi}.

\section{conclusions}

We have studied the  space-time symmetries and transformation properties of the non-commutative U(1) gauge theory, from the
dynamical point of view. The same analysis applies to other gauge groups.

We carried out our analysis by keeping an open view on the possible ways $\theta^{\mu \nu}$ could
transform. A possible scenario is the one with $\theta^{\mu \nu}$ behaving like a second rank tensor under Lorentz
transformations \cite{7A}. Another possibility is to see $\theta^{\mu \nu}$ as a back-ground metric, therefore reduce the
space-time transformations to the isometries of this metric which would leave the theory invariant.

>From our analysis it clearly follows that there is no reason for
$\theta^{\mu \nu}$ to transform under any space-time
transformation since the theory is not invariant under the
conformal transformations, with the only exception of
translations, and special choices of the parameters and/or of the
field configurations. This implies that $\theta^{\mu \nu}$ is an
invariant constant matrix in any frame of reference.

Since there is no invariant constant antisymmetric second rank tensor in four dimensions,
the physics necessarily depends on the frame of reference. Examples, for special choices of $\theta^{\mu \nu}$, are given in
\cite{Jac1} and \cite{Cai}, where it is shown that, in such a theory, the velocity of light depends on the direction of motion, like
in an anisotropic medium.

\acknowledgments We are grateful to R. Jackiw for introducing us
to the subject of symmetries in  non-commutative gauge theories,
and for very enjoyable discussions. A.I. has been partially
supported by the Fellowship ``Theoretical Physics of the
Fundamental Interactions'', University of Rome Tor Vergata. T.S.
was supported by the program ``Research Centres'' project n.
LN00A006 of the Ministry of Education, Youth and Sport of the
Czech Republic. This work was also supported in part through funds
provided by the U. S. Department of Energy (D.O.E) under
cooperative research agreement DF-FC02-94ER40818.


\begin{thebibliography}{99}

 \bibitem{Snyder} H. Snyder, Phys. Rev. 71 (1947), 38.

 \bibitem{Connes} A. Connes, "Non-Commutative Geometry", Academic Press, 1994.

 \bibitem{SW}     N. Seiberg, and E. Witten, JHEP 9909 (1999), 32.

 \bibitem{Jac1}   Z. Guralnik, R. Jackiw, S.Y. Pi, and A.P. Polychronakos,
          Phys. Lett. B 517 (2001), 450.

 \bibitem{Cai}    R. Cai, Phys. Lett. B 517 (2001), 457.

 \bibitem{Wess}   J. Madore, S. Schraml, P. Schupp, J. Wess,
          Eur. Phys. J. C 16 (2000), 161.

 \bibitem{7A}     A.A. Bichl, J.M. Grimstrup, H. Grosse, E. Kraus, L. Popp, M. Schweda, R. Wulkenhaar,
          {\it Noncommutative Lorentz Symmetry and the origin of the Seiberg-Witten Map},
          hep-th/0108045.

 \bibitem{Jac2} R. Jackiw, Acta Phys. Austr. Suppl. XXII (1980), 383.

\bibitem{Jac3} R. Jackiw, Phys. Rev. Lett. 41 (1978), 1635.

\bibitem{JacPi} R. Jackiw, and S.Y. Pi, {\it in preparation}.

 \bibitem{Io2}    A. Iorio, PhD Thesis, hep-th/0006198 and Phys. Lett. B 487 (2000), 171.\\
          A. Iorio, L. O'Raifeartaigh, S. Wolf,  Ann. Phys. 290 (2001), 156.
\bibitem{Lop}     J. Lopuszanski, ``An Introduction to Symmetry and Supersymmetry in Quantum Field Theory'',
          World Scientific, 1991.


 \bibitem{Io1}    A. Iorio, L. O'Raifeartaigh, I. Sachs, C. Wiesendanger,
          Nucl.Phys. B 495 (1997), 433.\\
          C.G. Callan, S. Coleman, and R. Jackiw, Ann. Phys. 59 (1970), 42.



%
 \end{thebibliography}
 \end{document}